\documentclass[submission, LectureNotes]{SciPost}

\usepackage{slashed}
\usepackage{bbold}

\newtheorem{exercise}{Exercise}

\allowdisplaybreaks

\begin{document}

\begin{center}{\Large \textbf{
Dark Matter Effective Theory
}}\end{center}

\begin{center}
Joachim Brod\textsuperscript{*}
\end{center}

\begin{center}
Department of Physics, University of Cincinnati, Cincinnati, OH 45221, USA
\\
* joachim.brod@uc.edu
\end{center}

\begin{center}
\today
\end{center}


\section*{Abstract}
{\bf Les Houches 2021 lectures on dark matter effective field theory
  (short course). The aim of these two lectures is to calculate the
  DM-nucleus cross section for a simple example, and then generalize
  to the treatment of general effective interactions of spin-1/2
  DM. Relativistic local operators, the heavy-DM effective theory, the
  chiral effective Lagrangian, and nuclear effective operators are
  briefly discussed.}

\vspace{10pt}
\noindent\rule{\textwidth}{1pt}
\tableofcontents\thispagestyle{fancy}
\noindent\rule{\textwidth}{1pt}
\vspace{10pt}

\section{Introduction}

In these two lectures on ``DM effective field theories'' we will
mainly be concerned with calculating the differential scattering rate
for DM scattering on nuclei:
\begin{equation}
  \frac{dR}{dq} = \frac{\rho_0}{m_A m_\text{DM}}
                  \int_{v_\text{min}} dv \, v f(v)
                  \frac{d\sigma}{dq}(v,q) \,.
\end{equation}
The left side of the equation can be integrated and compared with
experimental data. The right side is proportional to the local DM
density $\rho_0$; in our context, we regard it as given (astrophysical
input). $f(v)$ is the velocity distribution of DM in the halo; we also
regard it as given. For a given nuclear target $A$ with nucleus mass
$m_A$, we thus need to calculate the {\em differential cross section}
$d\sigma/dq$.

In general, calculating the cross section is a complicated problem. It
involves DM particles (whose properties like mass and spin are not
known) scattering on such complex objects as atomic nuclei -- strongly
bound states of quarks and gluons. One way to deal with this problem
is to split the calculation into many separate calculations
(``factorization'') that are simplified by making well-motivated
approximations (``power counting''). These are essential ideas of the
{\em effective field theory} approach.

In the first lecture, we will calculate in detail the DM scattering
cross section for a vector interaction ``from first principles'',
introducing some of the ideas of effective theory in passing. In the
second lecture, we will extend these ideas into a framework that works
also for more general interactions. Many advanced topics enter this
discussion; unfortunately, there is no time to properly explain all
these interesting concepts in these lectures. Instead, I will focus on
a qualitative understanding. However, for the benefit of the
interested reader, I list some introductory material (lectures,
reviews, and original articles) to the various topics here. The
selection is no means complete.

Introductions to the general ideas of effective field theories are
given by Pich~\cite{Pich:1998xt} and
Neubert~\cite{Neubert:2005mu}. Stewart has an
\href{https://ocw.mit.edu/courses/physics/8-851-effective-field-theory-spring-2013/index.htm}{online
  course} that is publically accessible. Another general review
article is the one by Georgi~\cite{Georgi:1993mps}.
Buras has written an introduction to the weak effective Hamiltonian in
his Les Houches lectures~\cite{Buras:1998raa}. The standard review by
Buchalla, Buras, and Lautenbacher~\cite{Buchalla:1995vs} is more
concise.
The lectures by Pich~\cite{Pich:1995bw} can serve as an introduction
to chiral perturbation theory. The book by Scherer and
Schindler~\cite{Scherer:2005ri} has many more details and discusses
also baryons. The book by Georgi~\cite{Georgi:1984zwz} also contains a
discussion of the chiral Lagrangian. Epelbaum discusses nuclear
physics based on the chiral approach in his lecture
notes~\cite{Epelbaum:2010nr}.
The heavy-DM effective theory was mainly adapted from the heavy-quark
effective theory as applied mainly in flavor physics. Manohar and Wise
have written a book on the topic~\cite{Manohar:2000dt}; two further
sets of lecture notes are by Neubert~\cite{Neubert:1993mb} and
Buchalla~\cite{Buchalla:2002pd}.


\section{Lecture 1: Calculation of a simple cross section}

To set the stage, we consider the kinematics of the elastic scattering
process. Let's assume we have a DM particle of mass $m_\chi =
100\,$GeV, scattering off an atomic xenon nucleus (mass roughly $m_A =
130\,$GeV). The escape velocity of our galaxy is about
$500\,$km/s,\footnote{Here, we neglect the motion of the earth; see
  Ref.~\cite{Lewin:1995rx} for a detailed discussion.} or in units of
speed of light $500\,\text{km/s}/(299792\,\text{km/s}) \sim
0.002$. This is the maximum speed of DM in the galactic halo -- DM can
safely be treated as nonrelativistic. We can use energy and momentum
conservation to estimate the momentum transfer in the scattering
process. It will be useful for later to find combinations that are
invariant under change of the coordinate system (as DM is
nonrelativistic, Galilean invariance will suffice). One obvious
candidate is the {\em momentum transfer}. Denoting the in- and
outgoing DM momenta by $\pmb{p}_1$, $\pmb{p}_2$ and the in- and
outgoing nuclear momenta by $\pmb{k}_1$, $\pmb{k}_2$, we define
$\pmb{q} \equiv \pmb{k}_2 - \pmb{k}_1 = \pmb{p}_1 -
\pmb{p}_2$. Another Galilean invariant quantity is the relative
incoming velocity, $\pmb{v} \equiv \pmb{v}_{\chi,\text{in}} -
\pmb{v}_{A,\text{in}}$.

Let's calculate the maximal momentum transfer in the lab frame,
scattering on a xenon nucleus at rest. We have $|\pmb{p}_1| = 0.002
m_\chi = 0.2\,$GeV, $|\pmb{k}_1| = 0$. The momentum transfer is
maximal for a ``head-on'' collision, such that $\pmb{p}_1$ and
$\pmb{p}_2$ are collinear. A straighforward calculation
gives\footnote{For collinear momenta, momentum conservation gives
  \begin{equation}\label{eq:qmax:mom}
    p_1 = k_2 \mp p_2 \quad \Rightarrow \quad p_2^2 = (p_1 - k_2)^2
  \end{equation}
  (the sign is negative if $m_\chi < m_A$, and positive
  otherwise). Energy conservation gives
  \begin{equation}\label{eq:qmax:en}
    \frac{p_2^2}{2m_\chi} + \frac{k_2^2}{2m_A} = \frac{p_1^2}{2m_\chi} 
    \Rightarrow
    p_2^2 = p_1^2 - \frac{m_\chi}{m_A} k_2^2 \,.
  \end{equation}
  Substituting Eq.~\eqref{eq:qmax:en} into Eq.~\eqref{eq:qmax:mom} and
  rearranging gives
  \begin{equation}
    2 p_1 = \bigg( 1 + \frac{m_\chi}{m_A} \bigg) k_2 \,,
  \end{equation}
  and solving for $k_2$ yields Eq.~\eqref{eq:qmax}.}
\begin{equation}\label{eq:qmax}
  |\pmb{q}|_\text{max} = \frac{2\mu_A}{m_\chi} |\pmb{p}_1| \,,
\end{equation}
where $\mu_A = m_A m_\chi/(m_A+m_\chi)$ is the nucleus-DM reduced
mass. In our numerical example, $|\pmb{q}|_\text{max} \sim
225\,$MeV. The maximal energy transferred to the nucleus is
\begin{equation}
  E_{A,\text{out,max}} = \frac{4m_A m_\chi}{(m_A+m_\chi)^2} E_{\chi,\text{in}} \,.
\end{equation}
Using $E_{\chi,\text{in}} = |\pmb{p}_1|^2/(2m_\chi)$, this gives
$E_{A,\text{out,max}} \sim 200\,$keV.

We see that energy and momentum transfer are very small compared to
the other scales in the problem, namely, the DM and nuclear masses. We
can use this to make some approximations that will simplify the
calculation of the cross section.

We did not yet talk about the force between DM and the nucleus. While
the scattering process itself is nonrelativistic, any realistic,
fundamental model of DM must be Lorentz invariant. In relativistic
quantum field theory, forces between particles are described by
exchanges of bosons. In the following, we consider a simple toy model
that describes fermionic DM interactions with SM quarks via the
exchange of a heavy vector particle (``a $Z'$ model''). The relevant
interaction Lagrangian\footnote{Strictly speaking, this is a
  Lagrangian density, such that the Lagrangian is given by $L = \int
  dx^4 {\mathcal L}$. To avoid clumsy language, I will follow common
  habit and do not distinguish the two where no confusion can arise.}
is
\begin{equation}
  {\mathcal L} = g_V V_\mu \bar\chi \gamma^\mu \chi
                 + g_V^\prime V_\mu \sum_{q=u,d} \bar q \gamma^\mu q \,.
\end{equation}
Here, $u$ and $d$ denote the up- and down-quark fields, while $V^\mu$
denotes the massive vector field. For simplicity, we assume equal
couplings $g_V'$ to up and down quarks, and that all other couplings
to SM particles vanish. The coupling between the vector particle and
DM is denoted by $g_V$. Using the Feynman rules corresponding to this
Lagrangian, we can calculate the transition amplitude ${\mathcal M}$
and then obtain the differential cross section for elastic scattering,
see Eq.~\eqref{eq:dsigmadER} below. It will be instructive to do this
calculation explicitly for our simple example.

\begin{figure}[t]
        \centering
        \includegraphics[width=0.3\textwidth]{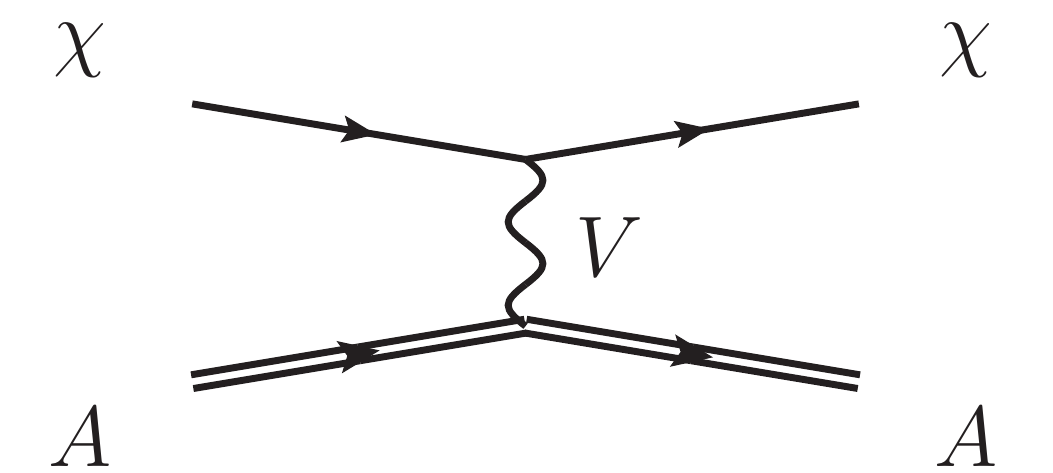}
        \caption{Leading order Feynman diagram for DM scattering on a
          nucleus via the exchange of a vector particle.
	\label{fig:LO}}
\end{figure}

At leading order, the transition amplitude is given in terms of the
S-matrix element by
\begin{equation}\label{eq:me:1}
\begin{split}
  S 
& = (2\pi)^4 i {\mathcal M} \delta^4 (p_\text{out} - p_\text{in}) \\
& = \frac{(-i)^2}{2!} (-ig_V) (-ig_V') \int d^4x d^4y
        \langle \chi(\pmb{p}_2,r') A(\pmb{k}_2,s') | \\ & \qquad \times
        T\{ \big(\bar\chi \slashed{V} \chi\big)(x) \sum_{q=u,d} \big(\bar q \slashed{V} q\big)(y)
            + (x\leftrightarrow y) \}
        | \chi(\pmb{p}_1,r) A(\pmb{k}_1,s) \rangle \,.
\end{split}
\end{equation}
The leading-order Feynman diagram for the scattering amplitude is
shown in Fig.~\ref{fig:LO}.

The DM current can be contracted with the external states as
usual. The vector fields can only be contracted internally and hence
yield a propagator factor
\begin{equation}
    D(x-y) = \langle 0 | T \{ V(x) V(y) \} | 0 \rangle
      = -i \int \frac{d^4k}{(2\pi)^4}
    \frac{g^{\mu\nu} - \frac{k^\mu k^\nu}{M_V^2}}{k^2 - M_V^2}
    e^{-ik(x-y)} \,.
\end{equation}
It is relatively straightforward to see that the ``gauge dependent''
piece proportional to $k^\mu k^\nu$ will not contribute to the
amplitude (at tree level, this follows essentially directly from
translational invariance), and we will drop this piece from now on.
Exchanging $x\leftrightarrow y$ in Eq.~\eqref{eq:me:1} cancels the
factor $1/2!$. Inserting the propagator gives
\begin{equation}
\begin{split}
  S & = -i g_V g_V' \int \frac{d^4k}{(2\pi)^4} \int d^4x d^4y
        \frac{g^{\mu\nu}}{k^2 - M_V^2}
        e^{-ik(x-y)} \\ & \hspace{6em} \times
        \langle \chi(\pmb{p}_2,r') A(\pmb{k}_2,s') | 
        \big(\bar\chi \gamma_\mu \chi\big)(x) \sum_{q=u,d} \big(\bar q \gamma_\nu q\big)(y)
        | \chi(\pmb{p}_1,r) A(\pmb{k}_1,s) \rangle \,.
\end{split}
\end{equation}
Now we need to consider the contractions of the external states with
the field operators. Since DM is completely neutral, the matrix
element in the second line factorizes into two independent parts:
\begin{equation}
\begin{split}
& \quad \langle \chi(\pmb{p}_2,r') A(\pmb{k}_2,s') | 
        \big(\bar\chi \gamma_\mu \chi\big)(x) \sum_{q=u,d} \big(\bar q \gamma_\nu q\big)(y)
        | \chi(\pmb{p}_1,r) A(\pmb{k}_1,s) \rangle \\
& = \langle \chi(\pmb{p}_2,r') | \big(\bar\chi \gamma_\mu \chi\big)(x) | \chi(\pmb{p}_1,r) \rangle
    \langle A(\pmb{k}_2,s') | \sum_{q=u,d} \big(\bar q \gamma_\nu q\big)(y) | A(\pmb{k}_1,s) \rangle \,.
\end{split}
\end{equation}

The ``DM factor'' is easy to evaluate, as the Lagrangian is written
directly in terms of DM fields. Application of the usual Feynman rules
(in this lecture, we follow the conventions of Peskin \&
Schroeder~\cite{Peskin:1995ev}) gives
\begin{equation}
  \langle \chi(\pmb{p}_2,r')|
  \big(\bar\chi \gamma^\mu \chi\big)(x)|
  \chi(\pmb{p}_1,r) \rangle
= \bar u (\pmb{p}_2,r') \gamma^\mu u (\pmb{p}_1,r) e^{i(p_2-p_1)\cdot x} \,.
\end{equation}
We will now take the nonrelativistic (NR) limit. Inserting the leading
expansion of the spinor function\footnote{We use the chiral
  representation and follow the conventions in
  Ref.~\cite{Peskin:1995ev}. The full solution for the Dirac spinor
  in these conventions is
  \begin{equation}
    u(\pmb{p}) = 
    \begin{pmatrix}
      \sqrt{p\cdot\sigma} \xi \\ \sqrt{p\cdot\bar\sigma} \xi
    \end{pmatrix}\,.
  \end{equation}
} in the NR limit, we find
\begin{equation}
  u (\pmb{p},r) = \sqrt{m_\chi} \begin{pmatrix}\xi\\\xi\end{pmatrix}_r \,.
\end{equation}
Here, $\xi$ is a NR two-component spinor for DM, with $(1,0)^T$
($(0,1)^T$) denoting spin up (down) along the $z$ axis. Recalling the
definition
\begin{equation}
  \gamma^\mu = \begin{pmatrix}0&\sigma^\mu\\\bar\sigma^\mu&0\end{pmatrix} \,,
\end{equation}
where $\sigma^\mu = (1,\sigma^i)$, $\bar\sigma^\mu = (1,-\sigma^i)$ in
terms of the usual Pauli matrices, we find
\begin{align}
\bar u (\pmb{p}_2,r') \gamma^0 u (\pmb{p}_1,r) & = 2m_\chi \delta_{r'r} \,,\\
\bar u (\pmb{p}_2,r') \gamma^i u (\pmb{p}_1,r) & = 0 \,.
\end{align}
Hence, the DM factor in the limit of small momentum transfer is just
\begin{equation}\label{eq:dmf}
  \langle \chi_\text{out}(\pmb{p}_2,r')|\big(\bar\chi \gamma^\mu \chi\big)(x) |\chi_\text{in}(\pmb{p}_1,r) \rangle
= 2 m_\chi \delta_0^\mu \delta_{r'r} e^{i(p_2-p_1)\cdot x} \,.
\end{equation}

The hadronic current $\langle A (\pmb{q}_2,s')| \big(\bar q \gamma_\mu
q \big) (y)| A (\pmb{q}_1,s) \rangle$ requires a bit more work. The
metric tensor in the propagator together with the Kronecker $\delta$
in Eq.~\eqref{eq:dmf} ensures that only the $\mu=0$ component will
contribute. There are two ways to proceed: One can show from the
explicit expressions of the quark fields that $\bar q \gamma^0 q
\propto a^\dagger a - b^\dagger b$ which just counts the number of
quarks minus the number of antiquarks. However, this calculation is
somewhat tedious. We can avoid it by following a different route that
is slightly more abstract but can be generalized later: we will use a
symmetry.  We recognize the integral over the zero component of the
quark bilinear as the conserved Noether charge, $Q_B$, of the baryon
current:
\begin{equation}
\begin{split}
  Q_B \equiv \int d^3 \pmb{y} \sum_{q=u,d} \bar q \gamma^0 q(y) \,.
\end{split}
\end{equation}
This just counts the number of baryons in the initial and final
state. (This is underlying the ``coherent enhancement'' of
spin-independent scattering.) In fact, the symmetry we used in this
case is exact, so we get an exact result. In general, we will use
approximate symmetries, which give approximate results.
\begin{exercise}
  Derive the Noether current for baryon number conservation. What is
  the corresponding symmetry of the SM Lagrangian?
\end{exercise}
We see that the hadronic part of the matrix elements just counts the
number of baryons minus the number of antibaryons in the nucleus
(recall that quarks have baryon number $+1/3$, while antiquarks have
baryon number $-1/3$). Using the results in App.~\ref{app:baryon}, the
``nucleus factor'' gives
\begin{equation}
\begin{split}
& \quad \langle A(\pmb{k}_2,s') | 
           \sum_{q=u,d} \big(\bar q \gamma^0 q\big)(y)
            | A(\pmb{k}_1,s) \rangle \\
& = \langle A(\pmb{k}_2,s') | 
           \sum_{q=u,d} \big(\bar q \gamma^0 q\big)(0) e^{i(k_2-k_1)\cdot y}
            | A(\pmb{k}_1,s) \rangle \\
& = e^{i(k_2-k_1)\cdot y} 2 k_1^0 A \delta_{s's} \,,
\end{split}
\end{equation}
where $A$ is the atomic number (number of nucleons) in the
nucleus. Note that in the NR limit, $k_1^0 = m_A$. Now we can combine
our results and find
\begin{equation}
\begin{split}
  S & = - 4A i g_V g_V' m_\chi m_A \delta_{r'r} \delta_{s's} \\
    & \qquad \times \int \frac{d^4k}{(2\pi)^4} \int d^4x d^4y
        \frac{1}{k^2 - M_V^2}
        e^{-ik(x-y)} e^{i(p_2-p_1)\cdot x} e^{i(k_2-k_1)\cdot y} \,.
\end{split}
\end{equation}
We can now easily perform the $x$ and $y$ integrals; they just yield
delta functions:
\begin{equation}
\begin{split}
  S & = - 4A i g_V g_V' m_\chi m_A \delta_{r'r} \delta_{s's} \\
    & \qquad \times \int \frac{d^4k}{(2\pi)^4}
        (2\pi)^4 \delta^4(p_2-p_1-k) (2\pi)^4 \delta^4(k+k_2-k_1)
        \frac{1}{k^2 - M_V^2} \,.
\end{split}
\end{equation}
The momentum integration is then also easy, and we obtain
\begin{equation}
  S = - 4A i g_V g_V' m_\chi m_A \delta_{r'r} \delta_{s's}
        (2\pi)^4 \delta^4(p_2+k_2-p_1-k_1)
        \frac{1}{q^2 - M_V^2} \,.
\end{equation}
The left-over delta function ensures four-momentum conservation in the
usual way. We have defined $q = k_2-k_1$ as above. We can now expand
in small $q$,
\begin{equation}
  \frac{1}{q^2 - M_V^2}
 = - \frac{1}{M_V^2} \times \frac{1}{1 - q^2/M_V^2}
 = - \frac{1}{M_V^2} \bigg[ 1 + {\mathcal O}\bigg( \frac{q^2}{M_V^2} \bigg) \bigg]\,,
\end{equation}
and retain only the leading term; this gives
\begin{equation}
  S = 4A i \frac{g_V g_V'}{M_V^2} m_\chi m_A \delta_{r'r} \delta_{s's}
        (2\pi)^4 \delta^4(p_2+k_2-p_1-k_1) \,.
\end{equation}
By definition, the transition matrix element is $S = (2\pi)^4 i
\delta^4(p_2+k_2-p_1-k_1) {\mathcal M}$, so
\begin{equation}
  {\mathcal M} = 4A g_V g_V' \frac{m_\chi m_A}{M_V^2} \delta_{r'r} \delta_{s's} \,.
\end{equation}

To calculate the cross section, we need $|{\mathcal M}|^2$, average
over initial spins, and sum over final spins (the spin components are
not measured in direct detection experiments). The average gives a
factor $1/2 \times 1/2 = 1/4$, and the sum over spins is easily
performed. We find
\begin{equation}
  \frac{1}{4} \sum_{rr'ss'} |{\mathcal M}|^2
= 16A^2 \big(g_V g_V'\big)^2 \frac{m_\chi^2 m_A^2}{M_V^4} \,,
\end{equation}
and the cross section is
\begin{equation}
  \sigma
= \frac{1}{\pi} \mu_{\chi A}^2 A^2 \frac{g_V^2 g_V^{\prime 2}}{M_V^4} \,,
\end{equation}
where we used $s = (E_{A,\text{in}} + E_{\chi,\text{in}})^2 = (m_A +
m_\chi)^2$ and introduce the reduced mass of the DM-nucleus system,
$\mu_{\chi A} = m_\chi m_A/(m_A + m_\chi)$.

This was straightforward\footnote{I should mention that we
  oversimplified the calculation a little bit. For large momentum
  transfer, the DM will be able to partially resolve the point
  nucleus, so we should take a ``form factor'' into account. See
  below.} but tedious! In the next lecture, we will generalize the
ideas used here and introduce appropriate {\em effective field
  theories}. The main idea is to perform the simplifications before we
start calculating! In this way, we can treat also harder examples.

\section{Lecture 2: Effective field theory for DM}

Now we will generalize the ideas of our toy model calculation into a
more general strategy, performing the simplifications from the
start. We need to generalize three different techniques that played a
role in the calculation above:
\begin{enumerate}
  \item Expansion of mediator propagator
  \item NR limit of DM currents
  \item Hadronic matrix elements of quark currents
\end{enumerate}
We will discuss these steps in turn.

\subsection*{Local DM Interactions}

The underlying idea of using effective field theory is to recognize
that typically several different {\em energy scales} contribute to a
given process. An energy scale is essentially any quantity with a mass
dimension. In our example above, the scales were the particle masses
$M_V$, $m_\chi$ and $m_A$, as well as the momentum transfer $q$. We
can then simplify the calculation by expanding in small dimensionless
ratios of these scales, and retaining only the leading terms. For
instance, above we expanded the propagator and kept the leading term
in $q^2/M_V^2$. The result was a contribution to the amplitude that is
the same as the one obtained from the {\em local operator}
\begin{equation}\label{eq:VV}
  Q_{1,q}^{(6)} = \big( \bar \chi \gamma^\mu \chi \big) \big( \bar q \gamma_\mu q \big) \,,
\end{equation}
with a coefficient $-g_V g_V' / M_V^2$. It is easy to imagine that
different types of mediators (vector, scalar, pseudoscalar, \ldots)
would yield different types of local interaction when expanded to
leading order in momentum transfer. Instead of ``integrating out'' all
kinds of different mediators explicitly, one can just write down all
possible local interactions. The ``rules of the game'' are to preserve
Lorentz invariance and conservation laws (such as of electric charge);
they manifest themselves as {\em symmetries}. Writing all such
operators that are linearly independent gives a {\em basis of
  operators}; the systematics is nicely explained in
Ref.~\cite{Grzadkowski:2010es}.  While there are infinitely many such
operators, there is only a finite number with a given {\em mass
  dimension} (see below). Instead of writing down the full basis, we
consider a second example,
\begin{equation}\label{eq:PP}
  Q_{8,q}^{(7)} = m_q \big( \bar \chi i \gamma_5 \chi \big) \big( \bar q i \gamma_5 q \big) \,.
\end{equation}
(A full basis up to mass dimension seven can be found, e.g., in
Refs.~\cite{Bishara:2016hek, Brod:2017bsw} whose numbering scheme we
followed here.) The number in the superscript denotes the mass
dimension of the operator. Since the Lagrangian density must have mass
dimension four (such that the Lagrangian is a dimensionless number),
it follows that the coefficient of any dimension-six operator must be
suppressed by an inverse square of a heavy mass (the vector mediator
mass in our example above); generally, we write this as
$1/\Lambda^2$. The dimension-seven operator is suppressed by
$1/\Lambda^3$. However, this is partially just a convention. We
included in the definition a factor of $m_q$ that arises in many
models with scalar or pseudoscalar exchange. So we could think of this
as $m_q/\Lambda \times 1/\Lambda^2$, where the first factor is related
to electroweak symmetry breaking, and the second factor arises from
integrating out the mediator.

In general, each of these operators comes with a ``Wilson''
coefficient whose value depends on the UV theory. We write the
Lagrangian density as
\begin{equation}\label{eq:lightDM:Lnf5}
{\mathcal L}_\chi=\sum_{a,d}
\hat C_{a}^{(d)} Q_a^{(d)}, 
\qquad {\rm where}\quad 
\hat C_{a}^{(d)}=\frac{C_{a}^{(d)}}{\Lambda^{d-4}}\,,
\end{equation}
summing over mass dimension $d$ and operators $a$. Our first example
above corresponds to $C_{1,q}^{(6)} = - g_V g_V'$ and $\Lambda = M_V$,
with all other Wilson coefficients zero.

\subsection*{Heavy DM Effective Theory}

If DM is nonrelativistic, its energy is dominated by its mass, and it
is useful to perform an expansion in powers of momentum divided by
mass. We treated a simple example in our explicit calculation
above. This example can be generalized as follows. Recall that the
(free) DM field $\chi(x)$ satisfies the Dirac equation,
\begin{equation}\label{eq:D.E.}
  (i\slashed \partial - m_\chi) \chi(x) = 0 \,.
\end{equation}
This can be interpreted as follows: via Fourier transformation,
$\partial_\mu$ corresponds to a four-momentum. The Dirac equation
implies that the energy contains the ``large mass''. In the NR regime
we are only interested in the kinetic energy part, since the mass does
not change. It is thus convenient to get rid of the implicit large
mass pieces in all energies. This can be done by splitting a general
DM four-momentum into $p^\mu = m_\chi v^\mu + k^\mu$, where $v^\mu$ is
the four-velocity, and the components of $k^\mu$ are small compared to
$m_\chi v^\mu$. We the split the DM field correspondingly:
\begin{equation}\label{eq:chi:split}
  \chi(x) = e^{-im_\chi v \cdot x} \big( \chi_v(x) + X_v(x) \big) \,,
\end{equation}
where
\begin{equation}
  \chi_v (x) =  e^{i m_\chi v \cdot x} \frac{1 + {\slashed v}}{2} \chi
  (x) \,, \qquad X_v (x) = e^{i m_\chi v \cdot x} \frac{1 - \slashed
    v}{2} \chi (x) \,,
\end{equation}
and we rescaled all fields by a factor $e^{i m_\chi v \cdot x}$. The
projectors $P_v^\pm = (1\pm\slashed{v})/2$ generalize the usual
decomposition of a spinor into ``large'' and ``small'' components in a
covariant way. Essentially, they project on the particles as opposed
to antiparticles, as the latter cannot be produced with
nonrelativistic energies. (In nonrelativistic QM this is known as a
``Foldy-Wouthuysen transformation''.) Due to the rescaling,
derivatives correspond to small energies and momenta. The field
$\chi_v(x)$ describes the NR DM modes, while $X_v(x)$ describes the
anti-particles modes. We will ``integrate them out'' as follows.
Multiplying Eq.~\eqref{eq:D.E.} by $(1-\slashed{v})/2$ yields
\begin{equation}
\begin{split}
                & \frac{1-\slashed{v}}{2} (i\slashed \partial - m_\chi) \chi(x) = 0 \\[1ex]
\Leftrightarrow \quad & \bigg( i\slashed \partial \frac{1+\slashed{v}}{2}
                  - i v \cdot \partial
                  - m_\chi \frac{1-\slashed{v}}{2} \bigg) 
                  e^{-im_\chi v \cdot x} \big( \chi_v(x) + X_v(x) \big) = 0 \\[1ex]
\Leftrightarrow \quad & e^{-im_\chi v \cdot x} 
                  \bigg( (i\slashed{\partial} + m_\chi \slashed{v}) \frac{1+\slashed{v}}{2}
                  - (i v \cdot \partial + m_\chi)
                  - m_\chi \frac{1-\slashed{v}}{2} \bigg) 
                  \big( \chi_v(x) + X_v(x) \big)
                  = 0 \\[1ex]
\Leftrightarrow \quad & (i\slashed{\partial} + m_\chi) \chi_v(x)
                  - (i v \cdot \partial + m_\chi)
                  \big( \chi_v(x) + X_v(x) \big)
                  = m_\chi X_v(x) \\[1ex]
\Leftrightarrow \quad & (i\slashed{\partial} - i v \cdot \partial) \chi_v(x)
                  - (i v \cdot \partial + 2 m_\chi) X_v(x)
                  = 0 \,,
\end{split}
\end{equation}
and thus
\begin{equation}\label{eq:eom:Xx}
  (i v \cdot \partial + 2 m_\chi) X_v(x) = i \slashed{\partial}_\perp \chi_v(x) \,.
\end{equation}
We used $(1+\slashed{v}) X_v = (1-\slashed{v}) \chi_v = 0$ and
$\slashed{v}\slashed \partial = - \slashed \partial\slashed{v} + 2 v
\cdot \partial$, and have defined $\partial_\perp^\mu \equiv
\partial^\mu - v^\mu v \cdot \partial$.  
Now we act with the inverse differential operators $(i v \cdot
\partial + 2 m_\chi)^{-1}$ on Eq.~\eqref{eq:eom:Xx} and obtain
\begin{equation}
  X_v(x) = \frac{i \slashed{\partial}_\perp}{i v \cdot \partial + 2 m_\chi} \chi_v(x) \,,
\end{equation}
or, inserting into Eq.~\eqref{eq:chi:split},
\begin{equation}\label{eq:chi:nr}
  \chi(x)
 = e^{-im_\chi v \cdot x}
   \bigg[ 1 + \frac{i \slashed{\partial}_\perp}{i v \cdot \partial + 2 m_\chi} \bigg]
   \chi_v(x) \,.
\end{equation}
We find the {\em Heavy Dark Matter Effective Theory (HDMET)}
Lagrangian by replacing the fields in the relativistic Lagrangian
using Eq.~\eqref{eq:chi:nr} and expanding the denominator in a power
series,
\begin{equation}
  \frac{1}{i v \cdot \partial + 2 m_\chi}
= \frac{1}{2 m_\chi} \Big[ 1 - \frac{i v \cdot \partial}{2 m_\chi} + \ldots \Big] \,.
\end{equation}
We will only write down the leading term of the Lagrangian:
\begin{equation} 
\label{eq:HDMET}
{\mathcal L}_\text{HDMET} = \bar \chi_v (i v \cdot \partial) \chi_v
 + \ldots + 
{\mathcal L}_{\chi_v} \,.
\end{equation}
\begin{exercise}
  Derive the leading term, starting from the relativistic Lagrangian
  $${\mathcal L} = \bar\chi (i\slashed{\partial} - m_\chi) \chi\,.$$
\end{exercise}
The term ${\mathcal L}_{\chi_v}$ contains the higher dimension
interaction operators. We will consider the two examples from above:
the vector current and the pseudoscalar current. Other currents can be
treated in the same way. We will derive only the leading terms in the
expansion. For the vector current we have
\begin{equation}
  \bar \chi \gamma^\mu \chi
  = \bar \chi_v e^{im_\chi v \cdot x} \gamma^\mu e^{-im_\chi v \cdot x} \chi_v
  = \bar \chi_v v^\mu \chi_v + \ldots \,.
\end{equation}
where we inserted Eq.~\eqref{eq:chi:nr} and used
\begin{equation}\label{eq:gamma:hdmet}
  \frac{1+\slashed{v}^\dagger}{2} \gamma^0 \gamma^\mu \frac{1+\slashed{v}}{2}
= \gamma^0 \frac{1+\slashed{v}}{2} \bigg[\frac{1 - \slashed{v}}{2} \gamma^\mu + v^\mu \bigg]
= \gamma^0 \frac{1+\slashed{v}}{2} v^\mu \,,
\end{equation}
and the ellipsis denotes higher-order terms. In the lab system, $v^\mu
= (1,0,0,0)$ and we recover our previous result of a contact
interaction.
For the pseudoscalar current, the momentum-independent term vanishes,
\begin{equation}
\begin{split}
  \frac{1+\slashed{v}^\dagger}{2} \gamma^0 \gamma_5 \frac{1+\slashed{v}}{2}
= \gamma^0 \frac{1+\slashed{v}}{2} \frac{1 - \slashed{v}}{2} \gamma_5
= 0 \,,
\end{split}
\end{equation}
so we need to go one order higher:
\begin{equation}
\begin{split}
  \bar \chi i\gamma_5 \chi
& = \bar \chi_v \Big[ 1 - \frac{i \stackrel{\leftarrow}{\slashed{\partial}}_\perp}{2m_\chi} \Big]
                e^{im_\chi v \cdot x} i\gamma_5 e^{-im_\chi v \cdot x} 
                \Big[ 1 + \frac{i \slashed{\partial}_\perp}{2m_\chi} \Big]
     \chi_v + \ldots \\
& = \bar \chi_v \Big[ 1 - \frac{i \stackrel{\leftarrow}{\slashed{\partial}}_\perp}{2m_\chi} \Big]
                i\gamma_5 
                \Big[ 1 + \frac{i \slashed{\partial}_\perp}{2m_\chi} \Big]
     \chi_v + \ldots \\
& = \frac{\partial_\mu}{m_\chi} \Big[ \bar \chi_v \frac{\gamma_\perp^\mu \gamma_5}{2} \chi_v \Big] + \ldots
  = \frac{\partial_\mu}{m_\chi} \big(\bar \chi_v S_\chi^\mu \chi_v \big) + \ldots \,,
\end{split}
\end{equation}
with $\gamma_\perp^\mu = \gamma^\mu - v^\mu v\cdot \gamma$. We see
that this interaction is ``momentum suppressed'' in the low-energy
limit.
\begin{exercise}
  Show the validity of the last equality in the heavy-DM limit. The
  relativistic generalization of the spin operator is defined as
  $S^\mu \equiv -\tfrac{1}{2} \epsilon^{\mu\nu\rho\lambda} J_{\nu\rho}
  v_\lambda$, with the sign convention $\epsilon^{0123} = +1$ for the
  Levi-Civita tensor, and $J^{\mu\nu} = \tfrac{1}{2} \sigma^{\mu\nu}$
  with $\sigma^{\mu\nu} = \tfrac{i}{2} [\gamma^\mu, \gamma^\nu]$ for
  spin 1/2. (This exercise is somewhat tedious.)
\end{exercise}

\subsection*{Chiral Effective Theory}

Passing to the NR limit was straightforward for the DM currents (as
long as we consider DM to be elementary). On the other hand, it is not
very useful to go to the NR limit for the quark fields, since QCD is
strongly coupled at low energies.

To obtain a physical amplitude, we really need to calculate a matrix
element between external nucleus states. The first step is to look at
the scattering on a single nucleon at a time (a justification will be
given later). We can express these single-nucleon matrix elements
exploiting all available symmetries in terms of so-called {\em form
  factors}. For instance, the general (elastic) vector-vector
interaction can be parameterized as
\begin{equation}
\label{vec:form:factor}
\langle N'|\bar q \gamma^\mu q|N\rangle
=\bar u_N'\Big[F_1^{q/N}(q^2)\gamma^\mu+\frac{i}{2m_N}F_2^{q/N}(q^2) \sigma^{\mu\nu}q_\nu\Big]u_N\,,
\end{equation}
where $F_{1,2}^{q/N}(q^2)$ are the form factors -- functions of the
four-momentum transfer $q^2$. We have seen above that, e.g.,
$F_{1}^{u/p}(0) = 2$ and $F_{1}^{d/p}(0) = 1$, etc.

Similarly form factors can be written down for other interactions
($V-A$, scalar, \ldots). However, the determination of the functions
$F(q^2)$ is not always as simple. For the electromagnetic vector
current, they can be measured in processes with photon exchange
(e.g. elastic electron-nucleon scattering). Some others can by
measured in neutrino scattering. Some, however, cannot (currently) be
measured.

A systematic approach is to exploit the {\em chiral symmetry of
  QCD}. I will give only the very basic idea of what is needed; see,
for instance, Refs.~\cite{Pich:1995bw, Scherer:2005ri, Georgi:1984zwz}
for details. The QCD Lagrangian for three massless quarks $q =
(u,d,s)$ is
\begin{equation}
  {\mathcal L}_\text{light quark}
= \bar q i \slashed{D} q
= \bar q_L i \slashed{D} q_L + \bar q_R i \slashed{D} q_R \,,
\end{equation}
where $D_\mu = \partial_\mu + ig_s T^a G_\mu^a$ is the covariant
derivative of QCD. It is invariant under {\em chiral rotations} of the
quark fields
\begin{equation}
  q_L \to L \, q_L \,, \qquad q_R\to R \, q_R \,,
\end{equation}
where $L \in SU(3)_L$, $R \in SU(3)_R$. 
\begin{exercise}
  Verify this explicitly. Show that quark mass terms break this
  symmetry.
\end{exercise}
At low energies, QCD is strongly coupled and the dynamics is (so far)
not analytically understood. However, we know that the chiral symmetry
$SU(3)_L \times SU(3)_R$ is spontaneously broken (in addition to the
explicit breaking by quark masses and QED effects). This is enough to
write down the most general effective theory of QCD in terms of its
low-energy degrees of freedom, the pions and nucleons. We collect the
pions into the matrix $U = \exp \big( i \sqrt{2} \Pi / f \big)$, where
\begin{equation}\label{eq:App:PiMat}
\Pi=
\begin{pmatrix}
\frac{\pi^0}{\sqrt2}+\frac{\eta_8}{\sqrt6} & \pi^+ &K^+ \\
\pi^- &  - \frac{\pi^0}{\sqrt2}+\frac{\eta_8}{\sqrt6} & K^0\\
K^- & \bar K^0 & -\frac{2\eta_8}{\sqrt6}
\end{pmatrix}
\end{equation} 
contains the Goldstone-boson fields, and $f = f_\pi \simeq 92\,$MeV
can be identified with the pion decay constant.  This matrix is
unitary and transforms as $U \to R U L^\dagger$ under $SU(3)_L \times
SU(3)_R$. We use it to construct an effective Lagrangian for the pion
fields that is invariant under these rotations. This Lagrangian will
be non-renormalizable, and the pion fields will transform nonlinearly
under axial rotations, reflecting the spontaneous breaking of the
symmetry. The leading-order (LO) term is simply
\begin{equation}
  {\mathcal L}_\text{ChPT,LO}
= \frac{f^2}{4} \text{Tr} \big( \partial_\mu U^\dagger \partial^\mu U \big) \,.
\end{equation}
Apart from a constant, there is no chirally invariant term without
derivatives. Effectively, the chiral Lagrangian is an expansion in
derivatives, or, in Fourier space and using our estimate above, in
small momenta. Chiral symmetry ensures that all terms are proportional
to the same factor $f^2$.

Explicit breaking effects can also be included. Consider the quark
masses. The mass term
\begin{equation}\label{eq:mass:partonic}
  {\mathcal L}_\text{quark mass}
= - \bar q M_q q
= - \bar q_L M_q q_R + \text{h.c.} \,,
\end{equation}
with $M_q = \text{diag}(m_u, m_d, m_s)$ is formally invariant under
chiral rotations if we let the quark mass matrix transform as $M_q \to
L M_q R^\dagger$. Recalling the transformation law $U \to
RUL^\dagger$, we see that the corresponding mass term in the effective
theory must be proportional to $\text{Tr} \big[ M_q U \big] +
\text{h.c.}$. Similar to Eq.~\eqref{eq:mass:partonic}, this term is
Hermitean and formally invariant under chiral rotations. Only its
coefficient cannot be predicted by chiral symmetry. Using the fact
that the quark mass matrix is actually real, the chiral Lagrangian
including the mass term is given by
\begin{equation}
  {\mathcal L}_\text{ChPT,LO}
= \frac{f^2}{4} \text{Tr} \big( \partial_\mu U^\dagger \partial^\mu U \big)
  + \frac{B_0 f^2}{2} \text{Tr} \big[ M_q \big( U + U^\dagger \big) \big] \,,
\end{equation}
where and $B_0 \sim 2.67\,$GeV is another low-energy constant that
cannot be predicted from symmetry arguments. (It is related to the
``quark condensate'' and can be determined, e.g., by lattice QCD.)
This Lagrangian can be used to show that $m_\pi^2 \propto m_q$ and
hence to extract the quark mass ratios from experimental data. The
procedure followed here of translating symmetry breaking terms from
the partonic to the effective Lagrangian is sometimes called ``spurion
method''.

In a similar way, DM matter interactions with pions can be written
down. For instance, let's consider our previous two examples -- the
couplings to pions of a DM vector current, $v^\mu \bar\chi_v \chi_v$,
coupled to a quark vector current, and a DM pseudoscalar current,
$\partial_\mu (\bar\chi_v S_\chi^\mu \chi_v)/m_\chi$, coupled to a
quark pseudoscalar current -- as in Eqs.~\eqref{eq:VV}
and~\eqref{eq:PP}. Treating the DM current as spurions like we did for
the quark mass terms, and keeping only leading terms, the chiral
Lagrangian can be shown to be\cite{Bishara:2016hek}
\begin{equation}\label{eq:ChPTLagrDM}
\begin{split}
{\mathcal L}_{\chi,{\rm ChPT}} \supset 
& - \frac{if^2}{2} 
    \text{Tr}\Big[\big(U (v\cdot \partial) U^\dagger+U^\dagger (v\cdot \partial) U\big)
                  \overline{C}_1^{(6)}
                  \bar\chi_v \chi_v \Big]\\ 
& - \frac{B_0 f^2}{2} \text{Tr}
    \Big[(U-U^\dagger) M_q \overline{C}_8^{(7)}
         \frac{i\partial_\mu}{m_\chi} (\bar\chi_v S_\chi^\mu \chi_v)\Big] \,,
\end{split}
\end{equation}
where $\overline{C}_i^{(d)} = \text{diag} \big(C_{i,u}^{(d)},
C_{i,d}^{(d)}, C_{i,s}^{(d)}\big)$. Note that these terms are
engineered to be Hermitian and even under parity (where $U \to
U^\dagger$). Expanding the $U$ matrices in inverse powers of $f$, it
is easy to see that the leading coupling of the vector current is to a
pair of pions, while the pseudoscalar current couples to a single
pion.
\begin{exercise}
  Perform the expansion explicitly to LO.
\end{exercise}

Our treatment of low-energy QCD for DM scattering is of course not
complete without the inclusion of nucleons. We will just quickly
summarize the results here; see, e.g., Ref.~\cite{Bishara:2016hek} for
details.

The first step is to pass to the low-energy limit also for
nucleons. This approximation is valid as long as $\pmb{q} \ll m_N$,
with $m_N \approx 1\,$GeV the nucleon mass. This is done by splitting
the baryon momentum $p^\mu$ into $p^\mu = m_N v^\mu + k^\mu$, with
$k^\mu$ the small residual momentum, and introducing a rescaled
nucleon field~\cite{Jenkins:1990jv}
\begin{equation}
  B_v(x) = \exp(im_N\slashed{v} v\cdot x) B(x) \,.
\end{equation}
The octet of baryon fields forms a $3\times 3$ matrix, given by
\begin{equation}
B_v=
\begin{pmatrix}
\frac{1}{\sqrt2} \Sigma_v^0+\frac{1}{\sqrt  6}\Lambda_v & \Sigma_v^+ & p_v\\
\Sigma_v^- & -\frac{1}{\sqrt2}\Sigma_v^0+\frac{1}{\sqrt 6}\Lambda_v & n_v \\
\Xi_v^- & \Xi_v^0 & -\frac{2}{\sqrt6}\Lambda_v
\end{pmatrix}\,.
\end{equation}
If we are interested only in tree-level processes, we can drop the
excited baryon states and just use
\begin{equation}
B_v=
\begin{pmatrix}
0 & 0 & p_v\\
0 & 0 & n_v \\
0 & 0 & 0
\end{pmatrix}\,.
\end{equation}
The interaction with the DM currents can be constructed using a
spurion method similar to what we did for the pions. See
Ref.~\cite{Bishara:2016hek} for details and general
expressions.\footnote{For baryons, there is considerable freedom in
  choosing the transformation under $SU(3)_L \times SU(3)_R$, as long
  as they transform as an octet under the unbroken ``vectorial''
  part. See Ref.~\cite{Georgi:1984zwz} for a nice discussion.} Here,
we will just collect the pieces that are relevant for the vector and
pseudoscalar currents, discussed above:
\begin{equation}
\begin{split}\label{eq:HBChPT0:expand}
{\cal L}_{\chi, {\rm HBChPT}}^{(0)}
 & \supset (\bar \chi_v \chi_v)(\bar p_v p_v) \Big( 2 \hat C_{1,u}^{(6,0)}+\hat C_{1,d}^{(6,0)} \Big)
+ (p_v \leftrightarrow n_v\,, u \leftrightarrow d)\,.
\end{split}
\end{equation}
Notice that this is essentially the same result as we obtain using the
baryon number conservation argument above -- no surprises here! Due to
parity conservation, there is no direct coupling of the pseudoscalar
current to a pure single-nucleon current; rather, the leading term has
the schematic form $\partial (\bar \chi_v S_\chi \chi_v) (\bar p_v
p_v) \pi + \ldots\,$. However, we have seen above that the
pseudoscalar current couples to a single pion. Hence we need the
pion-nucleon interaction,
\begin{equation}
{\cal L}_{\rm HBChPT}^{(1),{\rm QCD}}
\supset \frac{g_A}{f} \partial_\mu \big(\bar p_v S_N^\mu p_v - \bar n_v S_N^\mu n_v\big) \pi^0 \,,
\end{equation}
with the pion-nucleon coupling $g_A = 1.2756(13)$.

\begin{figure}
\begin{center}
\includegraphics[scale=1]{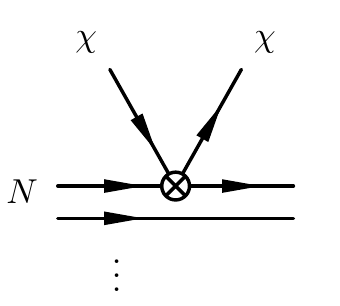}\hspace*{2cm}
\includegraphics[scale=1]{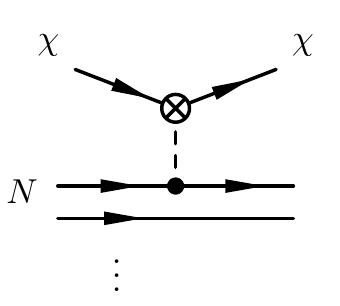}
\caption{Leading-order diagrams for DM-nucleus scattering. The
  effective DM--nucleon and DM--pion interactions are denoted by a
  circle, the dashed lines denote mesons, and the dots represent the
  remaining nucleon lines.}
         \label{fig:LOChPT}
\end{center}
\end{figure}

We now have all the ingredients to write Feynman diagrams for the
scattering, see Fig.~\ref{fig:LOChPT}. How do we decide which of these
diagrams are important? What about loops? Does the perturbative
expansion even make sense in the strongly-coupled regime? To answer
this question, recall that chiral perturbation theory is not a
coupling expansion, but a derivative expansion. Any chirally invariant
term without derivatives is just a constant and hence irrelevant. Now
consider a general amplitude involving an arbitrary number of pions,
carrying momenta at most of order $Q \ll m_N, m_\chi$. What is the
overall scaling $(Q/M)^\nu$ of the amplitude? Let's count: each pion
propagator contributes a factor $1/Q^2$. Each loop integral
contributes a factor $Q^4$. Each derivative in the interaction
vertices contributes a factor $Q$. So, for a diagram with $L$ loops,
$V_i$ interaction vertices with $d_i$ derivatives, and $I$ internal
lines, we find
\begin{equation}\label{eq:pre:power}
  \nu = \sum_i V_i d_i - 2I + 4L \,.
\end{equation}
Now it is known since Euler that in a given (connected) diagram, the
numbers of lines ($I$), vertices ($V_i$), and faces ($L$) is
constrained by the relation
\begin{equation}\label{eq:euler}
  L - I + \sum_i V_i = 1 \,.
\end{equation}
\begin{exercise}
  Can you find a proof for this relation?
\end{exercise}
Inserting the relation~\eqref{eq:euler} into Eq.~\eqref{eq:pre:power}
yields
\begin{equation}\label{eq:power:pion}
  \nu = \sum_i V_i (d_i-2) + 2L + 2 \,.
\end{equation}
Since $L \geq 0$, we see that the leading diagrams are tree-level
diagrams. Now this relation is strictly valid only for massless pions
with only self interactions. We now need to include pion masses, and
interaction with nucleons and DM. Since $Q^2$ will be at least of
order $m_\pi^2$, and $m_\pi^2 \sim m_q$, we should modify the power
counting relation to ($m_i$ are the factors of quark masses in vertex
$V_i$)
\begin{equation}\label{eq:power:pion:mass}
  \nu = \sum_i V_i (d_i+2m_i-2) + 2L + 2 \,.
\end{equation}
Next, we include nucleon. Since we treat them as nonrelativistic,
their propagators scale differently from the pion propagators:
\begin{equation}\label{eq:nuc:prop}
  \frac{-i(\slashed{p}+\slashed{q}+m_N)}{(p+q)^2 - m_N^2}
  \stackrel{q\to 0}{\longrightarrow}
  \frac{-i(\slashed{p}+m_N)}{2p\cdot q} \,,
\end{equation}
where we used $p^2 = m_N^2$. Hence, each nucleon propagator scales as
$1/Q$. Here, $p$ is the nucleon momentum and $q$ is the additional
momentum induced by pion interactions. (Note that the numerator in the
second term in Eq.~\eqref{eq:nuc:prop} is proportional to the
projector $P_v^+$.) Hence, a general Feynman diagram with $I_\pi$
internal pion lines and $I_N$ internal nucleon lines scales as
$Q^\nu$, where
\begin{equation}\label{eq:pre:power:nucleon}
  \nu = \sum_i V_i (d_i + 2m_i) - 2I_\pi - I_N + 4L + d_\chi \,.
\end{equation}
Here, I added the chiral dimension of the DM vertex (we will always
consider one single DM interaction for nuclear scattering). Now we can
again use the topological relation
\begin{equation}\label{eq:euler:2}
  L - I_\pi - I_N + \sum_i V_i = 1 \,,
\end{equation}
together with
\begin{equation}\label{eq:topo}
  2 I_N + E_N = \sum_i V_i n_i \,,
\end{equation}
where $n_i$ is the number of nucleon fields in the interaction $i$, to
obtain
\begin{equation}\label{eq:power:nucleon}
  \nu = \sum_i V_i \Big(d_i + 2m_i + \frac{n_i}{2} - 2\Big) + 2L - \frac{E_N}{2} + 2 + d_\chi \,.
\end{equation}
Finally, we note that the
relations~\eqref{eq:euler},~\eqref{eq:euler:2}~\eqref{eq:topo} are
valid for each connected component of the graph. Moreover, since the
nucleons are nonrelativistic, their number is conserved in the
scattering process. So we can sum over the connected components of the
graph and replace $E_N$ by two times the number of nucleon lines, $A$,
and obtain the final version of the power counting formula for a
$A$-nucleon irreducible graph with $C$ connected components:
\begin{equation}\label{eq:power:nucleon:final}
  \nu = 4 - A - 2C + 2L + \sum_i V_i \Big(d_i + 2m_i + \frac{n_i}{2} - 2\Big) + d_\chi \,.
\end{equation}

This power counting formula works well in our case where there a
single nucleon involved in the scattering. It tells us that the
leading contributions to the scattering amplitude come from tree-level
diagrams with single-nucleon interactions. Numerically, the
suppression is of order $30\%$ for each power of
momentum~\cite{Bishara:2016hek} -- taken to be somewhat larger than
expected in pure ChPT in order to account for additional effects of
the bound-state nucleons.

Going back to our two examples, the leading diagrams for DM with a
vector mediator and pseudoscalar mediator are given in
Fig.~\ref{fig:LOChPT}, left and right panel, respectively.

Finally, we need the nuclear matrix elements. The interactions are
given effectively in terms of nonrelativistic nucleon fields, possibly
contracted with spin matrices and derivatives. Again, one can write
down all possible allowed interaction that are now assumed to be
Galilean invariant. Here, we are only interested in interactions with
single nucleons and we can follow Ref.~\cite{Fitzpatrick:2012ix}. The
two operators we need are
\begin{equation}
{\mathcal O}_1^N = \mathbb{1}_\chi \mathbb{1}_N \,, \qquad
{\mathcal O}_6^N = \Big(\vec S_\chi \cdot \frac{\vec q}{m_N}\Big)
                   \, \Big(\vec S_N \cdot \frac{\vec q}{m_N}\Big) \,.
\end{equation}
The spin and unit matrices act on two-dimensional NR spinor space. The
NR field operators are not written explicitly due to some unfortunate
convention. More operators can be found in
Ref.~\cite{Fitzpatrick:2012ix}, and their connection to the UV DM
interactions is given in Ref.~\cite{Bishara:2016hek}. At first sight,
it might seems like the second operator is suppressed by $\vec
q^{\,\,2}/m_N^2$; recall, however, that the second diagram in
Fig.~\ref{fig:LOChPT} has a pion propagator that contributes a factor
$1/(\vec q^{\,\,2} + m_\pi^2)$, thus largely canceling the suppression
(see also the explicit result in Eq.~\eqref{eq:cNR:pseudoscalar}
below).

Fitzpatrick et al. have determined the nuclear matrix elements for a
variety of isotopes used in direct detection experiments (xenon,
germanium, fluorine, iodine, sodium) using shell model
calculations. The results are given numerically in
Ref.~\cite{Fitzpatrick:2012ix} in terms of nuclear form factors. They
can be used to calculate the NR differential cross
section~\cite{Anand:2013yka}:\footnote{To switch to the cross section
  differential in the momentum transfer, note that $q = \sqrt{2E_R
    m_A}$.}
\begin{equation}\label{eq:dsigmadER}
  \frac{d\sigma}{dE_R}
= \frac{m_A}{2\pi v^2}
  \frac{1}{2j_\chi+1} \frac{1}{2j_A+1} \sum_\text{spins} |{\mathcal M}|^2
\equiv \frac{m_A}{2\pi v^2} \sum_{ij} \sum_{N,N'=p,n} c_{i}^{N} c_{j}^{N'} F_{ij}^{(N,N')} \,.
\end{equation}
Here, the $c_i^N$ are the coefficients of the nuclear operators
${\mathcal O}_i^N$. For our two cases, the nuclear form factors are
given by $F_{1,1}^{(N,N')} = F_M^{(N,N')}$ and $F_{6,6}^{(N,N')} =
(\vec q\,{}^4/16) F_{\Sigma''}^{(N,N')}$, in the notation of
Ref.~\cite{Fitzpatrick:2012ix}.
For instance, for scattering on the most abundant xenon isotope
$^{132}$Xe:
\begin{equation}
  F_M^{(p,p)} = e^{-2y} \big( 2.9 - 11 y + 15 y^2 - 10 y^3 + 4 y^4 - 0.88 y^5 + 0.11 y^6 + \ldots \big) \times 10^{3} \,,
\end{equation}
while $F_{\Sigma''}^{(p,p)} = 0$ because $^{132}$Xe has spin
zero. Here, $b = \sqrt{41.467/(45A^{-1/3} - 25A^{-2/3})}\,$fm. The
(similar) form factors $F_M^{(p,n)}$ and $F_M^{(n,n)}$, as well as the
form factor for the other isotopes, can be found in
Ref.~\cite{Fitzpatrick:2012ix}. For a realistic cross section, we
should weight the xenon isotopes by their natural abundances.

It is instructive to also look at the scattering on fluorine,
$^{19}$F:
\begin{equation}
  F_M^{(p,p)} = e^{-2y} \big( 81 - 96 y + 36 y^2 - 4.7 y^3 + 0.19 y^4 \big) \,.
\end{equation}
Note that $2900/81 = 35.8 \sim (54/9)^2=36$, this is again the
coherent enhancement. Fluorine has nuclear spin $1/2$ and hence
\begin{equation}
  F_{\Sigma''}^{(p,p)} = e^{-2y} \big( 0.903 - 2.37 y + 2.35 y^2 - 1.05 y^3 + 0.175 y^4 \big)
\end{equation}
is non-zero.

It should now be clear how to calculate the general scattering cross
section. For our vector mediator example, we have
\begin{equation}
  c_1^p = 2 \hat C_{i,u}^{(6)} + \hat C_{i,d}^{(6)} \,, \quad
  c_1^n = \hat C_{i,u}^{(6)} + 2 \hat C_{i,d}^{(6)} \,,
\end{equation}
or, inserting the explicit values,
\begin{equation}
  c_1^p = c_1^n = - 3 \frac{g_V g_V'}{M_V^2} \,.
\end{equation}
Inserting this into Eq.~\eqref{eq:dsigmadER} immediately gives the
differential cross section. For the pseudoscalar interaction we find
\begin{equation}\label{eq:cNR:pseudoscalar}
\begin{split}
  c_6^p & = -\frac{B_0 m_N^2}{m_\chi} \frac{g_A}{m_\pi^2+\vec q\,{}^2}
           \big( m_u\,\hat C_{8,u}^{(7)}-m_d \,\hat C_{8,d}^{(7)}\big) \,, \\
  c_6^n & = \frac{B_0 m_N^2}{m_\chi} \frac{g_A}{m_\pi^2+\vec q\,{}^2}
           \big( m_u\,\hat C_{8,u}^{(7)}-m_d \,\hat C_{8,d}^{(7)}\big) \,.
\end{split}
\end{equation}
Note that the coefficients are {\em momentum dependent}. This has to
be taken into account when integrating the differential cross section
over the respective energy sensitivity windows for the different
experiments.

The whole chain of steps is quite straightforward even for more
general interactions, but tedious. Fortunately, public computer code
is available to perform these tasks. The program
\texttt{DirectDM}~\cite{Bishara:2017nnn} can be used to calculate the
coefficients of the nuclear operators given the UV interactions in
terms of Wilson coefficients. The code can be downloaded at
\begin{center}
\url{https://directdm.github.io}
\end{center}
Given the nuclear coefficients, \texttt{DMFormFactor} then allows for
the automatic calculation of the nuclear cross
section~\cite{Anand:2013yka}.

So far, we were concerned only with the leading approximation. Until
we have identified the precise nature of DM, this should be
sufficient. Nevertheless, we want to give an outlook on various
corrections that can occur. Some of the most widely studied
corrections are two-nucleon currents and perturbative radiative
corrections.

\begin{figure}
\includegraphics[scale=1]{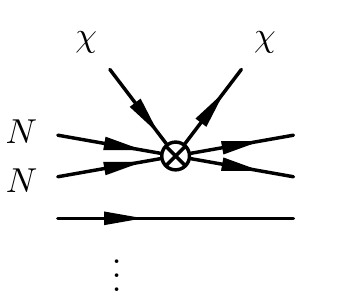}\hspace*{5mm}
\includegraphics[scale=1]{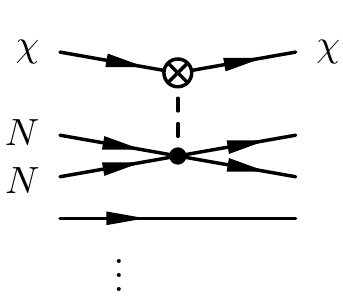}\hspace*{5mm}
\includegraphics[scale=1]{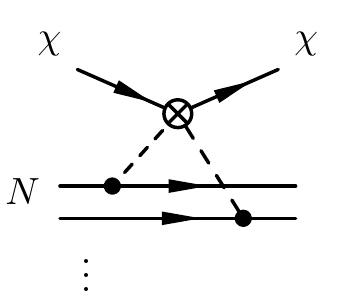}\hspace*{5mm}
\includegraphics[scale=1]{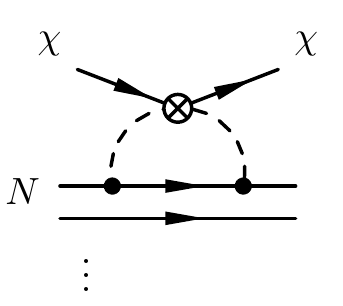}
\caption{Sample NLO diagrams for the DM-nucleon scattering inside
  nuclei. The effective DM--nucleon or DM--meson interaction is
  denoted by a box, the dashed lines denote pions.}
         \label{fig:NLOChPT}
\end{figure}

Some Feynman diagrams with higher-order nuclear contributions are
shown in Fig.~\ref{fig:NLOChPT}. The local two-nucleon interactions
(left two panels in Fig.~\ref{fig:NLOChPT}) are always suppressed by
three additional powers of momentum and are negligible. For certain
interactions, diagrams with loops or two single-nucleon interaction
(right two panels in Fig.~\ref{fig:NLOChPT}) may be suppressed by only
one power of momentum. This is the case for axialvector-vector,
scalar-scalar, and pseudoscalar-scalar interactions. Two-nucleon
currents can also be important if the leading contributions are absent
in specific models. More details and explicit results can be found in
Refs.~\cite{Cirigliano:2012pq, Hoferichter:2015ipa}.

\begin{figure}[t]\centering
\includegraphics[width=0.25\textwidth]{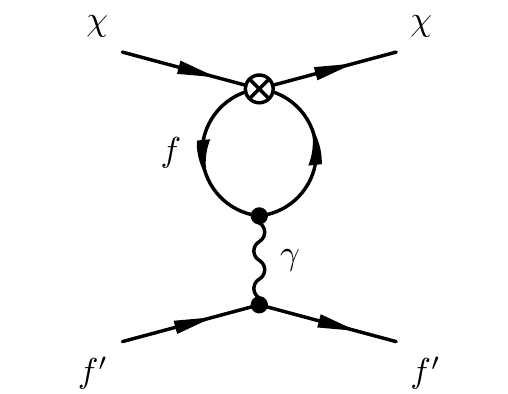}
\caption{The mixing of dimension-six four-fermion operators
         into each other via the photon penguin insertion.}
\label{fig:d6-photon-peng}
\end{figure}

Similarly, radiative corrections can have a large impact if models are
tuned such that leading contributions to nuclear scattering are absent
by construction. As a very simple example, consider DM with only {\em
  leptophilic} interactions -- DM couples only to leptons. At first
sight, it seems that scattering on atomic nuclei would be absent, but
single-photon exchange induces couplings to all
fermions~\cite{Kopp:2009et, DEramo:2017zqw} (see
Fig.~\ref{fig:d6-photon-peng}). More generally, the nuclear matrix
elements exhibit large hierarchies (spin-dependent
vs. spin-independent; momentum / velocity suppression). Whenever a
contribution to a large matrix element is not generated at leading
order in a model, but is not forbidden by a symmetry, it may be
generated via radiative corrections. For instance, the electroweak and
Yukawa interactions of the SM break parity, and loop-induced
contributions to nuclear scattering can be larger than the leading
terms by orders of magnitude (see Refs.~\cite{Crivellin:2014qxa,
  DEramo:2014nmf, DEramo:2016gos} for an example with top quarks). The
most important effects for fermionic DM are included in the
\texttt{DirectDM} code~\cite{Bishara:2018vix}.

\section{Conclusion}

We have seen that using effective field theory is very suitable in
describing the nonrelativistic scattering process of dark matter on
atomic nuclei. It allows us to calculate the event rates for a wide
class of dark matter interactions in term of UV parameters and a
handful of low-energy constants. Frequently, ``Dark Matter Effective
Field Theory'' refers specifically to the formalism developed in
Ref.~\cite{Fitzpatrick:2012ix}. In my opinion, this viewpoint is too
restricted and has a number of drawbacks:\footnote{I should hasten to
  point out that many of these issues have already been mentioned in
  Ref.~\cite{Fitzpatrick:2012ix}.} (a) Dark matter models are
typically formulated in terms of elementary particle fields (quarks,
leptons, gluons, \ldots), not in terms of nucleon fields. The
connection between these fundamental interactions (summarized in terms
of local operators and Wilson coefficients) should be made
explicit. (b) The coefficients of the nuclear operators are often
assumed to be constant, which is frequently not the case (see, for
instance, Eq.~\eqref{eq:cNR:pseudoscalar}). This becomes manifest if
one considers the whole tower of effective theories appropriate for
dark matter direct detection. (c) Radiative corrections can have a
large impact on the event rate in certain circumstances. These
corrections need to be calculated in a partonic effective theory.

The practical take-home message is the following. The most convenient
``meeting point'' between the hadronic physics and dark-matter model
building (for instance, when presenting experimental results in terms
of global fits) seems to be the effective theory in terms of local
operators defined at a scale $\mu = 2\,$GeV, just above the chiral
symmetry breaking scale of QCD. The connection to lower-scale physics
can then be obtained via the chiral approach described in these
lectures. The connection to realistic dark-matter models is equally
straightforward, employing the usual perturbative techniques of
matching and renormalization-group evolution, thus capturing the
leading radiative corrections~\cite{Bishara:2018vix}.

\section*{Acknowledgements}

The author would like to thank Le\'{o}n Manuel de la Vega, as well as
the anonymous referees of the lecture notes, for very valuable
comments on the manuscript.

\appendix

\section{Baryon number conservation}\label{app:baryon}

We are interested in the general form of matrix elements of the baryon
current
\begin{equation}
  {\mathcal J}_B^\mu(x) = \sum_q \bar \psi_q \gamma^\mu \psi_q \,.
\end{equation}
Translational invariance implies
\begin{equation}\label{eq:JQED:translation}
  \langle \pmb{p}', \sigma' | {\mathcal J}^\mu(x) | \pmb{p}, \sigma \rangle
= e^{-i(p-p')\cdot x} 
  \langle \pmb{p}', \sigma' | {\mathcal J}^\mu(0) | \pmb{p}, \sigma \rangle \,.
\end{equation}
(This can be derived as follows. In position space, the momentum
operator acts as
\begin{equation}
  [P_\mu, O(x)] = -i \frac{\partial}{\partial x^\mu} O(x) \,,
\end{equation}
on any local field operator $O(x)$. It follows
\begin{equation}
  \langle \beta | [P_\mu, O(x)] | \alpha \rangle
  = (p_\beta - p_\alpha)_\mu \langle \beta | O(x) | \alpha \rangle
  = -i \frac{\partial}{\partial x^\mu} \langle \beta | O(x) | \alpha \rangle \,,
\end{equation}
and so
\begin{equation}
  \langle \beta | O(x) | \alpha \rangle = \exp[i(p_\beta - p_\alpha)
    \cdot x] \langle \beta | O(0) | \alpha \rangle \,,
\end{equation}
which proves Eq.~\eqref{eq:JQED:translation}.) Setting $\mu = 0$ in
Eq.~\eqref{eq:JQED:translation} and integrating over $\pmb{x}$ gives
\begin{equation}
  \langle \pmb{p}', \sigma' | Q_B | \pmb{p}, \sigma \rangle
= (2\pi)^3 \delta^3(\pmb{p}' - \pmb{p})
  \langle \pmb{p}', \sigma' | {\mathcal J}^0(0) | \pmb{p}, \sigma \rangle \,,
\end{equation}
where we used the definition of (baryon) charge $Q_B \equiv \int
d^3\pmb{x} {\mathcal J}^0 (x)$. Denoting the charge of the state $|
\pmb{p}, \sigma \rangle$ by $B$, we find
\begin{equation}\label{eq:J0:ME}
  \langle \pmb{p}, \sigma' | {\mathcal J}^0(0) | \pmb{p}, \sigma \rangle
= 2 p^0 B \delta_{\sigma'\sigma} \,,
\end{equation}
where $p^0 \equiv \sqrt{\pmb{p}^2 + m^2}$.

\section{Solutions to exercises}

\subsection*{Exercise 1}

The Lagrangian
\begin{equation}
  {\mathcal L}_\text{quark}
= \bar q i \slashed{D} q - \bar q M_q q
\end{equation}
is invariant under the phase transformation $q \to e^{i\epsilon} q$,
with $\epsilon$ real. Infinitesimally, $q \to q + i\epsilon
q$. Noether's theorem tells us that for an infinitesimal symmetry
$\psi \to \psi + i\epsilon {\mathcal F}$, the current
\begin{equation}
  {\mathcal J}^\mu
= -i \frac{\partial {\mathcal L}}{\partial (\partial^\mu \psi)} {\mathcal F}
\end{equation}
is conserved. This gives
\begin{equation}
  {\mathcal J}_B^\mu
= \bar q \gamma^\mu q \,.
\end{equation}

\subsection*{Exercise 2}

Inserting Eq.~\eqref{eq:chi:nr} into the Lagrangian and keeping only
the leading terms gives
\begin{equation}
\begin{split}
    \bar\chi (i\slashed{\partial} - m_\chi) \chi
& \to \bar\chi_v e^{im_\chi v \cdot x} (i\slashed{\partial} - m_\chi) e^{-im_\chi v \cdot x} \chi_v \\
& =   \bar\chi_v (i\slashed{\partial} + m_\chi \slashed{v} - m_\chi) \chi_v
  =   \bar\chi_v (iv \cdot \partial) \chi_v \,,
\end{split}
\end{equation}
where we used $\slashed{v} \chi_v = \chi_v$ in the second-to-last and
Eq.~\eqref{eq:gamma:hdmet} in the last step.

\subsection*{Exercise 3}

Let us first check that $S^\mu$ is indeed the spin operator. We work
in the rest frame, $v=(1,0,0,0)$, and using the chiral representation
of the Dirac matrices:
\begin{equation}
\begin{split}
S^i & = - \frac{1}{2} \epsilon^{ijk0} J_{jk} = \frac{1}{2}
\epsilon^{0ijk} \frac{i}{4} [\gamma_j, \gamma_k] = \frac{i}{8}
\epsilon^{0ijk} [\gamma^j, \gamma^k] = \frac{i}{8} \epsilon^{ijk} 2i
\epsilon^{kjl} 
\begin{pmatrix}
\sigma^l&0\\0&\sigma^l
\end{pmatrix}\\
& = \frac{1}{2} \delta^{il}
\begin{pmatrix}
\sigma^l&0\\0&\sigma^l
\end{pmatrix} \,,
\end{split}
\end{equation}
or
\begin{equation}
\vec S = \frac{1}{2}
\begin{pmatrix}
\vec\sigma&0\\0&\vec\sigma
\end{pmatrix} \,.
\end{equation}
We define $\gamma_5 = i\gamma^0\gamma^1\gamma^2\gamma^3 \equiv
-\tfrac{i}{4!}  \epsilon^{\mu\nu\rho\sigma} \gamma_\mu \gamma_\nu
\gamma_\rho \gamma_\sigma$. From this you can show that
$\sigma^{\mu\nu} \gamma_5 = i \epsilon^{\mu\nu\rho\sigma}
\sigma_{\rho\sigma}/2$. Hence we can show that
\begin{equation}
S^\mu = \frac{i}{2} \sigma^{\mu\sigma} \gamma_5 v_\sigma \,,
\end{equation}
so that in the rest frame (using again the chiral representation)
\begin{equation}
\vec S = \frac{1}{2} 
\begin{pmatrix}
\vec \sigma&0\\0&\vec \sigma
\end{pmatrix}\,,
\end{equation}
as before. At last, we note that between NR spinors we can use the
projector $P_v^+ = (1+\slashed{v})/2$. We have
\begin{equation}
[P_v, \gamma^\mu]
= \frac{1+\slashed{v}}{2} \gamma^\mu - \gamma^\mu \frac{1+\slashed{v}}{2}
= v^\mu + \gamma^\mu \frac{1-\slashed{v}}{2} - \gamma^\mu \frac{1+\slashed{v}}{2}
= v^\mu - \gamma^\mu \slashed{v}
\end{equation}
We can use this to show
\begin{equation}
  P_v^+ S^\mu P_v^+
= \frac{i}{2} P_v^+ \sigma^{\mu\sigma} P_v^- \gamma_5 v_\sigma 
= \frac{i}{2} P_v^+ \frac{i}{2} (2v^\mu v\cdot \gamma - 2 \gamma^\mu) \gamma_5 P_v
= \frac{1}{2} \gamma_\perp^\mu \gamma_5\,.
\end{equation}
We have used $P_v^+ P_v^- = 0$ and $\slashed{v} P_v^- = - P_v^-$ in
intermediate steps.

\subsection*{Exercise 4}

The light-quark Lagrangian is
\begin{equation}
  {\mathcal L}_\text{light quark}
= \bar q_L i \slashed{D} q_L + \bar q_R i \slashed{D} q_R \,.
\end{equation}
The first term transforms as
\begin{equation}
  \bar q_L i \slashed{D} q_L
  \to \bar q_L L^\dagger i \slashed{D} L q_L
= \bar q_L i \slashed{D} q_L \,,
\end{equation}
and similarly for the second term. The quark mass term transforms as
\begin{equation}
  - \bar q_L M_q q_R + \text{h.c.}
  \to - \bar q_L L^\dagger M_q R q_R + \text{h.c.} \,.
\end{equation}
Note that the quark mass term is invariant for $L=R$ if all quark
masses are equal.

\subsection*{Exercise 5}

We have $U = \exp \big( i \sqrt{2} \Pi / f \big)$, so expanding the
exponential
\begin{equation}
  U = 1 + i \frac{\sqrt{2} \Pi}{f} + \ldots \,,
\end{equation}
we see that
\begin{equation}
\begin{split}
&  U (v \cdot \partial) U^\dagger + U^\dagger (v \cdot \partial) U \\
& = \bigg(1 + i \frac{\sqrt{2} \Pi}{f} \bigg)
    \bigg(- i \frac{\sqrt{2} (v \cdot \partial)\Pi}{f} \bigg) + 
    \bigg(1 - i \frac{\sqrt{2} \Pi}{f} \bigg)
    \bigg(i \frac{\sqrt{2} (v \cdot \partial)\Pi}{f} \bigg) + \ldots \\
& = 2 \frac{\Pi (v \cdot \partial)\Pi}{f^2} + \ldots \,.
\end{split}
\end{equation}
quadratic in pion fields, and
\begin{equation}
  U - U^\dagger = i \frac{2\sqrt{2} \Pi}{f} + \ldots\,,
\end{equation}
linear in pion fields.

\subsection*{Exercise 6}

Try mathematical induction.

\bibliography{references}

\begin{thebibliography}{10}
\providecommand{\url}[1]{\texttt{#1}}
\providecommand{\urlprefix}{URL }
\expandafter\ifx\csname urlstyle\endcsname\relax
  \providecommand{\doi}[1]{doi:\discretionary{}{}{}#1}\else
  \providecommand{\doi}{doi:\discretionary{}{}{}\begingroup
  \urlstyle{rm}\Url}\fi
\providecommand{\eprint}[2][]{\url{https://arxiv.org/abs/#2}}

\bibitem{Pich:1998xt}
A.~Pich,
\newblock \emph{{Effective field theory: Course}},
\newblock In \emph{{Les Houches Summer School in Theoretical Physics, Session
  68: Probing the Standard Model of Particle Interactions}} (1998),
  \eprint{hep-ph/9806303}.

\bibitem{Neubert:2005mu}
M.~Neubert,
\newblock \emph{{Effective field theory and heavy quark physics}},
\newblock In \emph{{Theoretical Advanced Study Institute in Elementary Particle
  Physics}: {Physics in D $\geq$ 4}},
\newblock \doi{10.1142/9789812773579_0004} (2005), \eprint{hep-ph/0512222}.

\bibitem{Georgi:1993mps}
H.~Georgi,
\newblock \emph{{Effective field theory}},
\newblock Ann. Rev. Nucl. Part. Sci. \textbf{43}, 209 (1993),
\newblock \doi{10.1146/annurev.ns.43.120193.001233}.

\bibitem{Buras:1998raa}
A.~J. Buras,
\newblock \emph{{Weak Hamiltonian, CP violation and rare decays}},
\newblock In \emph{{Les Houches Summer School in Theoretical Physics, Session
  68: Probing the Standard Model of Particle Interactions}} (1998),
  \eprint{hep-ph/9806471}.

\bibitem{Buchalla:1995vs}
G.~Buchalla, A.~J. Buras and M.~E. Lautenbacher,
\newblock \emph{{Weak decays beyond leading logarithms}},
\newblock Rev. Mod. Phys. \textbf{68}, 1125 (1996),
\newblock \doi{10.1103/RevModPhys.68.1125},
\newblock \eprint{hep-ph/9512380}.

\bibitem{Pich:1995bw}
A.~Pich,
\newblock \emph{{Chiral perturbation theory}},
\newblock Rept. Prog. Phys. \textbf{58}, 563 (1995),
\newblock \doi{10.1088/0034-4885/58/6/001},
\newblock \eprint{hep-ph/9502366}.

\bibitem{Scherer:2005ri}
S.~Scherer and M.~R. Schindler,
\newblock \emph{{A Chiral perturbation theory primer}}  (2005),
\newblock \eprint{hep-ph/0505265}.

\bibitem{Georgi:1984zwz}
H.~Georgi,
\newblock \emph{{Weak Interactions and Modern Particle Theory}},
\newblock ISBN 978-0-8053-3163-9 (1984).

\bibitem{Epelbaum:2010nr}
E.~Epelbaum,
\newblock \emph{{Nuclear Forces from Chiral Effective Field Theory: A Primer}}
  (2010), \eprint{1001.3229}.

\bibitem{Manohar:2000dt}
A.~V. Manohar and M.~B. Wise,
\newblock \emph{{Heavy quark physics}}, vol.~10,
\newblock ISBN 978-0-521-03757-0 (2000).

\bibitem{Neubert:1993mb}
M.~Neubert,
\newblock \emph{{Heavy quark symmetry}},
\newblock Phys. Rept. \textbf{245}, 259 (1994),
\newblock \doi{10.1016/0370-1573(94)90091-4},
\newblock \eprint{hep-ph/9306320}.

\bibitem{Buchalla:2002pd}
G.~Buchalla,
\newblock \emph{{Heavy quark theory}},
\newblock In \emph{{55th Scottish Universities Summer School in Physics: Heavy
  Flavor Physics (SUSSP 2001)}} (2002), \eprint{hep-ph/0202092}.

\bibitem{Lewin:1995rx}
J.~D. Lewin and P.~F. Smith,
\newblock \emph{{Review of mathematics, numerical factors, and corrections for
  dark matter experiments based on elastic nuclear recoil}},
\newblock Astropart. Phys. \textbf{6}, 87 (1996),
\newblock \doi{10.1016/S0927-6505(96)00047-3}.

\bibitem{Peskin:1995ev}
M.~E. Peskin and D.~V. Schroeder,
\newblock \emph{{An Introduction to quantum field theory}},
\newblock Addison-Wesley, Reading, USA,
\newblock ISBN 978-0-201-50397-5 (1995).

\bibitem{Grzadkowski:2010es}
B.~Grzadkowski, M.~Iskrzynski, M.~Misiak and J.~Rosiek,
\newblock \emph{{Dimension-Six Terms in the Standard Model Lagrangian}},
\newblock JHEP \textbf{10}, 085 (2010),
\newblock \doi{10.1007/JHEP10(2010)085},
\newblock \eprint{1008.4884}.

\bibitem{Bishara:2016hek}
F.~Bishara, J.~Brod, B.~Grinstein and J.~Zupan,
\newblock \emph{{Chiral Effective Theory of Dark Matter Direct Detection}},
\newblock JCAP \textbf{02}, 009 (2017),
\newblock \doi{10.1088/1475-7516/2017/02/009},
\newblock \eprint{1611.00368}.

\bibitem{Brod:2017bsw}
J.~Brod, A.~Gootjes-Dreesbach, M.~Tammaro and J.~Zupan,
\newblock \emph{{Effective Field Theory for Dark Matter Direct Detection up to
  Dimension Seven}},
\newblock JHEP \textbf{10}, 065 (2018),
\newblock \doi{10.1007/JHEP10(2018)065},
\newblock \eprint{1710.10218}.

\bibitem{Jenkins:1990jv}
E.~E. Jenkins and A.~V. Manohar,
\newblock \emph{{Baryon chiral perturbation theory using a heavy fermion
  Lagrangian}},
\newblock Phys. Lett. B \textbf{255}, 558 (1991),
\newblock \doi{10.1016/0370-2693(91)90266-S}.

\bibitem{Fitzpatrick:2012ix}
A.~L. Fitzpatrick, W.~Haxton, E.~Katz, N.~Lubbers and Y.~Xu,
\newblock \emph{{The Effective Field Theory of Dark Matter Direct Detection}},
\newblock JCAP \textbf{02}, 004 (2013),
\newblock \doi{10.1088/1475-7516/2013/02/004},
\newblock \eprint{1203.3542}.

\bibitem{Anand:2013yka}
N.~Anand, A.~L. Fitzpatrick and W.~C. Haxton,
\newblock \emph{{Weakly interacting massive particle-nucleus elastic scattering
  response}},
\newblock Phys. Rev. C \textbf{89}(6), 065501 (2014),
\newblock \doi{10.1103/PhysRevC.89.065501},
\newblock \eprint{1308.6288}.

\bibitem{Bishara:2017nnn}
F.~Bishara, J.~Brod, B.~Grinstein and J.~Zupan,
\newblock \emph{{DirectDM: a tool for dark matter direct detection}}  (2017),
\newblock \eprint{1708.02678}.

\bibitem{Cirigliano:2012pq}
V.~Cirigliano, M.~L. Graesser and G.~Ovanesyan,
\newblock \emph{{WIMP-nucleus scattering in chiral effective theory}},
\newblock JHEP \textbf{10}, 025 (2012),
\newblock \doi{10.1007/JHEP10(2012)025},
\newblock \eprint{1205.2695}.

\bibitem{Hoferichter:2015ipa}
M.~Hoferichter, P.~Klos and A.~Schwenk,
\newblock \emph{{Chiral power counting of one- and two-body currents in direct
  detection of dark matter}},
\newblock Phys. Lett. B \textbf{746}, 410 (2015),
\newblock \doi{10.1016/j.physletb.2015.05.041},
\newblock \eprint{1503.04811}.

\bibitem{Kopp:2009et}
J.~Kopp, V.~Niro, T.~Schwetz and J.~Zupan,
\newblock \emph{{DAMA/LIBRA and leptonically interacting Dark Matter}},
\newblock Phys. Rev. D \textbf{80}, 083502 (2009),
\newblock \doi{10.1103/PhysRevD.80.083502},
\newblock \eprint{0907.3159}.

\bibitem{DEramo:2017zqw}
F.~D'Eramo, B.~J. Kavanagh and P.~Panci,
\newblock \emph{{Probing Leptophilic Dark Sectors with Hadronic Processes}},
\newblock Phys. Lett. B \textbf{771}, 339 (2017),
\newblock \doi{10.1016/j.physletb.2017.05.063},
\newblock \eprint{1702.00016}.

\bibitem{Crivellin:2014qxa}
A.~Crivellin, F.~D'Eramo and M.~Procura,
\newblock \emph{{New Constraints on Dark Matter Effective Theories from
  Standard Model Loops}},
\newblock Phys. Rev. Lett. \textbf{112}, 191304 (2014),
\newblock \doi{10.1103/PhysRevLett.112.191304},
\newblock \eprint{1402.1173}.

\bibitem{DEramo:2014nmf}
F.~D'Eramo and M.~Procura,
\newblock \emph{{Connecting Dark Matter UV Complete Models to Direct Detection
  Rates via Effective Field Theory}},
\newblock JHEP \textbf{04}, 054 (2015),
\newblock \doi{10.1007/JHEP04(2015)054},
\newblock \eprint{1411.3342}.

\bibitem{DEramo:2016gos}
F.~D'Eramo, B.~J. Kavanagh and P.~Panci,
\newblock \emph{{You can hide but you have to run: direct detection with vector
  mediators}},
\newblock JHEP \textbf{08}, 111 (2016),
\newblock \doi{10.1007/JHEP08(2016)111},
\newblock \eprint{1605.04917}.

\bibitem{Bishara:2018vix}
F.~Bishara, J.~Brod, B.~Grinstein and J.~Zupan,
\newblock \emph{{Renormalization Group Effects in Dark Matter Interactions}},
\newblock JHEP \textbf{03}, 089 (2020),
\newblock \doi{10.1007/JHEP03(2020)089},
\newblock \eprint{1809.03506}.

\end{thebibliography}

\nolinenumbers

\end{document}